\documentclass[aip,chaos,reprint]{revtex4-1}

\usepackage{graphicx}
\usepackage{amssymb,amsfonts,amsmath}
\usepackage{epsf}
\usepackage{subfigure}
\usepackage{epstopdf}
\usepackage{dcolumn}
\usepackage{bm}
\usepackage{mathrsfs}
\usepackage{mhchem}
\usepackage{color}
\usepackage{ulem}

\newcommand{\be}{\begin{equation}}
\newcommand{\ee}{\end{equation}}
\newcommand{\bea}{\begin{eqnarray}}
\newcommand{\eea}{\end{eqnarray}}

\begin{document}

\title{\bf Stochastic approach to entropy production in chemical chaos}

\author{Pierre Gaspard}
\affiliation{Center for Nonlinear Phenomena and Complex Systems,\\
Universit\'e Libre de Bruxelles (U.L.B.), Code Postal 231, Campus Plaine,
B-1050 Brussels, Belgium}

\begin{abstract}
Methods are presented to evaluate the entropy production rate in stochastic reactive systems.  These methods are shown to be consistent with known results from nonequilibrium chemical thermodynamics.  Moreover, it is proved that the time average of the entropy production rate can be decomposed into the contributions of the cycles obtained from the stoichiometric matrix in both stochastic processes and deterministic systems.  These methods are applied to a complex reaction network constructed on the basis of R\"ossler's reinjection principle and featuring chemical chaos.
\end{abstract}

\maketitle

{\bf Far from equilibrium, autocatalysis may generate self-sustained oscillations in chemical reaction networks such as the Belousov-Zhabotinsky system.  The oscillations may be periodic, quasiperiodic, or chaotic.  In this latter case, they are referred to as chemical chaos or chemical turbulence, since pioneering work by Otto R\"ossler leading to the discovery of this phenomenon.  Such reactive systems are driven away from equilibrium with the free energy of reactants supplied from outside, so that energy is dissipated and entropy produced inside.  In chemical reaction networks, some reactions may be driven by the other ones in the direction opposite to spontaneous dissipation, leading to chemical energy conversion like in engines.  Moreover, molecular fluctuations are known to manifest themselves in small systems, where time evolution is ruled by stochastic processes.  For such fluctuating chemical reaction networks, the evaluation of thermodynamic properties remains today a challenging problem.}

\section{Introduction}

The theoretical prediction of chemical chaos, also called chemical turbulence, and its experimental discovery owe much to the pioneering contributions of Otto R\"ossler.  In 1976, he found chaotic behavior by numerical integration in a four-dimensional dynamical system modeling reaction and transport by diffusion between two compartments.\cite{R76ZN1168}  He also invented a series of prototype chaotic systems in dimension three,\cite{R76PLA,R76ZN1664,R77S,R79NY} which is the minimal dimension where chaotic behavior may exist for coupled first-order ordinary differential equations, as well as hyperchaos in dimension four.\cite{R79NY,R79PLA}  Inspired by the geometry of flows in three-dimensional phase spaces, R\"ossler introduced the {\it reinjection principle}.\cite{R76ZN259}  According to this principle, chaotic behavior may be induced if the dynamics evolves on a sigmoidal slow manifold with motion spiraling out of its lower branch, jumping to its upper branch with reinjection back onto the lower branch.

Moreover, in 1978, R\"ossler cosigned with Wegmann a paper reporting experimental evidence for chaos in the Belousov-Zhabotinsky reaction.\cite{RW78}  This paper confirmed the prediction of chaos in far-from-equilibrium chemical reactions\cite{R76ZN1168,R73} and triggered a surge in the experimental study of chemical chaos.\cite{HHM79,HM81,TRMS81,SWS82,S83,E83,R83,RSS83,BPV84,AAR87,S91,EP98}

On the side of theory, R\"ossler proposed with Willamowski in 1980 a chemical reaction network obeying the mass action law and generating chaotic behavior in a three-dimensional phase space.\cite{WR80,AC88} Later on, other chemical reaction networks with chemical chaos were proposed.\cite{GN83,SGW89}  These models are important because they provide a mechanistic understanding of chemical chaos in terms of colliding and reacting particles.  In this regard, such reaction networks play a pivotal role because they can be used to define Markovian stochastic processes for modeling chemical reactions at the mesoscopic level of description where molecular fluctuations manifest themselves.  In particular, stochastic processes based on the Willamowski-R\"ossler chemical reaction network have been analyzed to investigate the effects of molecular fluctuations on chemical chaos.\cite{WK93,GM93,GN93,GB96,VD16}  These investigations have shown that the stationary probability distribution of the stochastic process in phase space typically has a coarse structure looking alike the invariant probability density of the corresponding chaotic attractor in the weak noise limit.  Furthermore, the entropy production rate has been studied for the deterministic Willamowski-R\"ossler chemical reaction network.\cite{GNR96,SU05}

Entropy production is the keystone of nonequilibrium thermodynamics.\cite{D36,P67,N79,GM84,KP98}  In reaction networks driven far from equilibrium by the supply of free energy from reactant species,
entropy is produced due to energy dissipation by the chemical reactions.  Since several reactions are often coupled together in the network, such systems can perform energy conversion from one form to another like in  engines.\cite{RRR81}  Here, the form of energy is the free energy stored in either one or another of the reactant or product species.  Such chemical energy conversions can be characterized by their rate and their contribution to entropy production in the framework of nonequilibrium thermodynamics.  In small reactive systems where molecular fluctuations manifest themselves, similar issues can be addressed within stochastic thermodynamics, which is a research field pioneered in the mid-eighties by Nicolis, Van~den~Broeck, and coworkers.\cite{LVN84,MLN86,V86}  Recent advances in stochastic thermodynamics have shown how to systematically establish the balances of energy and entropy in many kinds of stochastic physicochemical systems.\cite{LS99,G04,AG04,AG07,AG08JCP,AG08PRE,TS08,S05,S11,S12,EV10,EV10I,EV10II,E12,V13,VE15,H15,D15,DDN16,XHX08,XHX09,XH10,RXH11,XZ18}

The purpose of this paper is to show that the methods of stochastic thermodynamics can be applied to complex chemical reaction networks featuring chemical chaos.  The vehicle of the present study is the reversible version of the chemical reaction network proposed in Ref.~\onlinecite{GN83}.  The inclusion of the reversed reactions is essential in order to investigate thermodynamic properties because the entropy production rate would be infinite if the reversed reactions had vanishing rates.  The chemical reaction network of Ref.~\onlinecite{GN83} is constructed using R\"ossler's reinjection principle and its deterministic dynamics has chaotic attractors in a three-dimensional phase space.  Shil'nikov conditions for homoclinic chaos are satisfied for a codimension-one subset in parameter space.\cite{S65}  The reaction network is composed of ten reactions and so many reversed ones, several of them being trimolecular.  The entropy production rate evaluated by the methods of stochastic thermodynamics is shown to agree with the values given by standard thermodynamics.  Furthermore, it is proved that, in reactive systems obeying the mass action law, the entropy production rate can be decomposed into different contributions corresponding to the cycles of the stoichiometric matrix.  In the studied reaction network, the chemical conversion of one species into another is shown to change its direction, which is characterized in terms of the thermodynamic efficiency of the conversion.  These thermodynamic properties are obtained in the deterministic and stochastic approaches with consistent values between the two.

The paper is organized as follows.  Section~\ref{Sec:stoch-react} summarizes the stochastic approach to reactive systems.  The method to obtain the entropy production rate in this approach is presented in Sec.~\ref{Sec:entr-prod}.  The results obtained in the deterministic limit are given in Sec.~\ref{Sec:det-lim}.  In this section, it is shown that the time average of the entropy production rate can be decomposed into the affinities and the rates associated with cycles of the stoichiometric matrix.  Furthermore, another method is inferred to compute the entropy production rate in the stochastic approach, which is equivalent to the one presented in Sec.~\ref{Sec:entr-prod}. In Sec.~\ref{Sec:GN83}, the results are applied to the chemical reaction network proposed in Ref.~\onlinecite{GN83}.  Section~\ref{Conclusion} draws the conclusion.

\section{Stochastic approach to reactive systems}
\label{Sec:stoch-react}

\subsection{Network of elementary reactions}

The following chemical reaction network is considered:
\be
\sum_{i=1}^e \tilde\nu_{i \rho}^{(+)}\; {\rm A}_i + \sum_{j=1}^s \nu_{j \rho}^{(+)}\; {\rm X}_j 
\ \underset{W_{-\rho}}{\overset{W_{+\rho}}{\rightleftharpoons}} \ 
\sum_{i=1}^e \tilde\nu_{i \rho}^{(-)}\; {\rm A}_i + \sum_{j=1}^s \nu_{j \rho}^{(-)}\; {\rm X}_j
\label{reactions}
\ee
with $\rho=1,2,\dots,r$, where $\{ {\rm A}_i\}_{i=1}^e$ are reactant or product molecular species, $\{ {\rm X}_j\}_{j=1}^s$ are intermediate molecular species, while $\tilde\nu_{i \rho}^{(\pm)}$ and $\nu_{j \rho}^{(\pm)}$ are non-negative integers giving the respective numbers of  molecules that are incoming or outgoing the reaction $\rho$.  The concentrations of these species inside the volume $V$ of the system are respectively denoted $\{ a_i\}_{i=1}^e$ and $\{ x_j\}_{j=1}^s$.  The system is supposed to be well mixed so that these concentrations are uniform on average across the volume $V$.

The reason for making the distinction between the intermediate and other species is that it is often assumed in the literature~\cite{EP98,KP98,LVN84,Schl71,Schl72,NP71,N72,NP77} that some species $\{ {\rm A}_i\}_{i=1}^e$ have their concentrations held constant during the reactive process.  This is achieved by maintaining these species in excess with respect to the intermediate species: $a_i\gg x_j$.  The species in excess are said to be chemostatted since their chemical potentials remain constant in time.\cite{PE14,RE16,RE18a,RE18b}   The reaction network is driven out of equilibrium if the concentrations of the chemostatted species are not in their equilibrium Guldberg-Waage ratios.\cite{D36}  In this regard, these species are playing the roles of reactant and product for the process.  The reactant species should be supplied from outside and the product species evacuated from inside in order to maintain constant their concentrations.  In particular, this assumption is made for enzymatic reactions in biochemical kinetics, where the substrate and product species are often supposed to have fixed concentrations.\cite{S75,WRE18}
 
In contrast, the molecular numbers $\pmb{X}(t)\equiv\{ X_j(t)\}_{j=1}^s\in{\mathbb N}^s$ of the intermediate species inside the system are evolving in time, because reactive events are randomly occurring at the transition rates $W_{\pm\rho}$ for the forward and backward reactions~(\ref{reactions}), respectively.  Hence, the concentrations $x_j(t)\equiv X_j(t)/V$ of the intermediate species are also varying in time.

Even though the reactant and product species have their concentrations $\{ a_i\}_{i=1}^e$ held fixed inside the system, these species should be supplied from or evacuated to the exterior of the system, so that the numbers $\pmb{A}(t)=\{ A_i(t)\}_{i=1}^e$ of the reactant and product species that are consumed or produced are changing in time.  Actually, these numbers are drifting in time, as long as the reaction network is maintained in a nonequilibrium regime.

\subsection{Stochastic reactive process}

The time evolution of the molecular numbers $\pmb{A}(t)$ and $\pmb{X}(t)$ can be expressed in terms of the numbers $\pmb{n}(t)\equiv \{ n_{\rho}(t) \}_{\rho=\pm 1}^{\pm r}$ of forward and backward reactive events that have occurred since the beginning of the process at time $t=0$.  These counters allow us to determine the fluxes of matter in the reaction network~(\ref{reactions}) as well as the composition of the system at every instant of time.  Introducing the stoichiometric coefficients of the different species according to
\be
\tilde\nu_{i \rho}\equiv \tilde\nu_{i \rho}^{(-)}-\tilde\nu_{i \rho}^{(+)} \qquad\mbox{and}\qquad  \nu_{j \rho}\equiv \nu_{j \rho}^{(-)}-\nu_{j \rho}^{(+)} \, ,
\label{stoichio}
\ee
the molecular numbers can be written at time $t$ as
\bea
A_i(t) &=& A_i(0) +\sum_{\rho=1}^{r} \tilde\nu_{i \rho} \left[ n_{+\rho}(t)-n_{-\rho}(t)\right] \nonumber\\
&& \qquad\qquad\qquad (i=1,2,\dots,e)\, , \label{DA}\\
X_j(t) &=& X_j(0) +\sum_{\rho=1}^{r} \nu_{j \rho} \left[ n_{+\rho}(t)-n_{-\rho}(t)\right] \nonumber\\
&& \qquad\qquad\qquad (j=1,2,\dots,s)\, , \label{DX}
\eea
in terms of the random variables $n_{\pm \rho}(t)$ counting the reactive events of the forward and backward elementary reactions $\rho=1,2,\dots,r$.  

Equation~(\ref{DX}) shows that we need at least as many observed intermediate species $\{{\rm X}_j\}$ as there are reactions, in order to determine the signed cumulated numbers $N_{\rho}\equiv n_{+\rho}-n_{-\rho}$ of elementary reactions $\rho=1,2,\dots,r$ from the observation of the intermediate species.  Actually, this inference is possible if $s\ge r$ and if the matrix $(\nu_{j \rho})$ of stoichiometric coefficients for $j=1,2,\dots,s$ and $\rho=1,2,\dots,r$ has full rank (i.e., its rank is equal to the number $r$ of elementary reactions), so that the signed cumulated numbers $\{N_{\rho}(t)\}_{\rho=1}^{r}$ of reactive events can be determined from the molecular numbers $\{X_j(t)-X_j(0)\}_{j=1}^s$.  However, if $s<r$, there are too few intermediate species to determine the signed cumulated numbers $\{N_{\rho}(t)\}_{\rho=1}^r$ of elementary reactions.  In such cases, some of the species $\{{\rm A}_i\}$ should also be observed to perform this determination.

The transition rates are denoted as follows,
\be\label{rates}
W_{\rho}(\pmb{X'}\vert\pmb{X})\quad\mbox{for the transition}\quad
\pmb{X} {\overset{\rho}\longrightarrow}\, \pmb{X'}=\pmb{X}+ \pmb{\nu}_{\rho} \, ,
\ee
where $\rho=\pm 1,\pm 2,\dots,\pm r$, $\pmb{X}=\{ X_j\}_{j=1}^s$, and $\pmb{\nu}_{\rho}=\{\nu_{j \rho}\}_{j=1}^s$ are the stoichiometric coefficients~(\ref{stoichio}).\cite{Gardiner}

According to the {\it mass action law} of chemical kinetics, the transition rates of the forward reactions are given by\cite{LVN84,NP71,N72,NP77}
\be\label{rates_mass_action}
W_{\rho}(\pmb{X}+ \pmb{\nu}_{\rho}\vert \pmb{X})= V \, k_{\rho}\; \prod_{i=1}^e a_i^{\tilde\nu_{i \rho}^{(+)} } \;  \prod_{j=1}^s \prod_{n=1}^{\nu_{j \rho}^{(+)}} \frac{X_j-n+1}{V}
\ee
in terms of the concentrations $\{ a_i\}_{i=1}^e$ of reactant or product species, the integers $\tilde\nu_{i \rho}^{(\pm)}$ and $\nu_{j \rho}^{(\pm)}$ defined in Eq.~(\ref{reactions}), and the molecular numbers $\{ X_j\}_{j=1}^s$ of the intermediate species involved in the reaction.  Similar expressions hold for the transition rates of the backward reactions.

With these transition rates, we can define a Markovian stochastic process for the time evolution of the joint random variables $\pmb{X}(t)$ and $\pmb{n}(t)$.  Gillespie's algorithm provides an exact Monte Carlo method for the numerical simulation of such Markovian stochastic processes.\cite{G76,G77}

\subsection{Chemical master equations}

If $\Delta\pmb{n}_{\rho}=-\Delta\pmb{n}_{-\rho}$ denotes the jump of the counters $\pmb{n}(t)$ upon the reactive event $\rho$, the master equation ruling the time evolution of the joint probability distribution $P(\pmb{X},\pmb{n},t)$ of the reactive process is given by
\bea
&& \frac{d}{dt} P(\pmb{X},\pmb{n},t) \nonumber\\
&& = \sum_{\rho=\pm 1}^{\pm r} \Big[ W_{\rho}(\pmb{X}\vert\pmb{X}-\pmb{\nu}_{\rho})\,
P(\pmb{X}-\pmb{\nu}_{\rho},\pmb{n}-\Delta\pmb{n}_{\rho},t) \nonumber\\
&&\qquad\qquad - W_{-\rho}(\pmb{X}-\pmb{\nu}_{\rho}\vert\pmb{X})\,
P(\pmb{X},\pmb{n},t)\Big]
\label{full.master.eq}
\eea
with the stoichiometric coefficients $\pmb{\nu}_{\rho}=-\pmb{\nu}_{-\rho}$.\cite{AG08PRE}

Summing over the counters $\pmb{n}$, we obtain the standard chemical master equation\cite{NP71,N72,NP77,Gardiner,S76,vK81}
\bea
&& \frac{d}{dt} {\mathscr P}(\pmb{X},t) = \sum_{\rho=\pm 1}^{\pm r} \Big[ W_{\rho}(\pmb{X}\vert\pmb{X}-\pmb{\nu}_{\rho})\, {\mathscr P}(\pmb{X}-\pmb{\nu}_{\rho},t)\nonumber\\
&&\qquad\qquad\qquad\qquad - W_{-\rho}(\pmb{X}-\pmb{\nu}_{\rho}\vert\pmb{X})\,
{\mathscr P}(\pmb{X},t)\Big]
\label{master.eq}
\eea
for the marginal probability distribution of the molecular numbers of the intermediate species
\be
{\mathscr P}(\pmb{X},t) = \sum_{\pmb{n}} P(\pmb{X},\pmb{n},t) \, .
\label{P(X)}
\ee
We note that the probability distribution~(\ref{P(X)}) of the intermediate species may reached a stationary distribution ${\mathscr P}_{\rm st}(\pmb{X})$, which justifies the terminology of {\it nonequilibrium steady state}.

In general, the counters $\pmb{n}(t)$ undergo random walks with positive diffusivities according to the central limit theorem.  Moreover, if the process is driven out of equilibrium, these variables are randomly drifting with non-vanishing rates, even if the molecular numbers $\pmb{X}(t)$ of the intermediate species follow a stationary process.  Their random drifts are known to obey a multivariate fluctuation theorem.\cite{AG07,G13NJP}  
If the reservoirs of reactant and product species are arbitrarily large, the counters $n_{\pm\rho}$ can be defined modulo ${\mathscr N}$ with ${\mathscr N}\gg 1$.  In this case, the stationary probability distribution of the stochastic process is given by $P_{\rm st}(\pmb{X},\pmb{n})={\mathscr P}_{\rm st}(\pmb{X})/{\mathscr N}^{2r}$.

\subsection{Random paths and their probability distribution}
\label{Subsec:rndm-paths}

The random paths of the process ruled by the master equation~(\ref{full.master.eq}) can be denoted
\be
{\cal X}(t) = (\pmb{X}_0,\pmb{n}_0)\to (\pmb{X}_1,\pmb{n}_1)
\to \cdots\to(\pmb{X}_m,\pmb{n}_m) \; ,
\label{path+n}
\ee
or, more shortly,
\be
{\cal X}(t) = \pmb{X}_0\; {\overset{\rho_1}\longrightarrow} \; \pmb{X}_1
\; {\overset{\rho_2}\longrightarrow} \; \cdots\;
{\overset{\rho_m}\longrightarrow}\; \pmb{X}_m \; ,
\label{path}
\ee
where $\{\rho_1,\rho_2,\dots,\rho_m\}$ is the sequence of reactive events occurring along the path.  If the reactive event $\rho$ occurs, the corresponding counter is incremented by unity according to $n_{\rho}{\overset{\rho}\to}\, n_{\rho}+1$ (for $\rho=\pm 1,\pm 2,\dots,\pm r$), so that Eqs.~(\ref{path+n}) and~(\ref{path}) provide equivalent descriptions of the path.  

The probability of the random paths~(\ref{path+n}) or~(\ref{path}) can be obtained by considering their stroboscopic observations every time step $\tau$.  Since the process is Markovian, the probability that the random variables take the values $(\pmb{X}_l,\pmb{n}_l)$ at time $t_l=t_0+l\tau$ while the initial values are sampled according to their stationary distribution is given by
\be\label{path-prob}
{\mathsf P}_{\rm st}[{\cal X}(t)] = \prod_{l=1}^n P_{\tau}(\pmb{X}_l,\pmb{n}_l\vert \pmb{X}_{l-1},\pmb{n}_{l-1}) \, P_{\rm st}(\pmb{X}_0,\pmb{n}_0)
\ee
in terms of the conditional probability to find the values $(\pmb{X}_l,\pmb{n}_l)$ at time $t_l=t_{l-1}+\tau$ given that the values were equal to $(\pmb{X}_{l-1},\pmb{n}_{l-1})$ at time $t_{l-1}$.   For small enough sampling time~$\tau$, these conditional probabilities can be expressed as
\be
P_{\tau}(\pmb{X'},\pmb{n'}\vert \pmb{X},\pmb{n}) = \left\{
\begin{array}{l}
W_{\rho}(\pmb{X'}\vert\pmb{X}) \, \tau + O(\tau^2) \, ,\\
\quad\mbox{if} \ n'_{\rho}=n_{\rho}+1 \, , \\
\quad\mbox{and}\ n'_{\tilde\rho}\ne n_{\tilde\rho} \ ,\forall \tilde\rho\ne \rho\, ; \\ \\
1-\gamma(\pmb{X}) \, \tau + O(\tau^2) \, , \\
\quad\mbox{if} \ n'_{\rho}=n_{\rho} \, , \forall \rho \, ;
\end{array}
\right.
\ee
with the following rate of escape from the state $\pmb{X}$,
\be\label{esc_rate}
\gamma(\pmb{X}) \equiv \sum_{\rho=\pm 1}^{\pm r} W_{\rho}(\pmb{X}+\pmb{\nu}_{\rho}\vert\pmb{X}) \, .
\ee
If the transitions $\pmb{X}_k=\pmb{X}_{k-1}+\pmb{\nu}_{\rho_k}$ occur at the successive times $t_1<t_2<\cdots<t_k<\cdots<t_m$, the path probability~(\ref{path-prob}) is thus obtained in the limit $\tau\to 0$ as\cite{vK81}
\bea\label{path-prob-2}
&&{\mathsf P}_{\rm st}[{\cal X}(t)] = {\rm e}^{-\gamma(\pmb{X}_m)(t-t_m)} \nonumber\\
&&\times \prod_{k=1}^m \left[ W_{\rho_k}(\pmb{X}_k\vert\pmb{X}_{k-1}) \, \tau \, {\rm e}^{-\gamma(\pmb{X}_{k-1})(t_k-t_{k-1})} \right] \nonumber\\
&&\times P_{\rm st}(\pmb{X}_0,\pmb{n}_0) \, .
\eea

We note that the time reversal of the path~(\ref{path}) can be written as
\be
{\cal X}^{\rm R}(t) =\pmb{X}_m\; {\overset{-\rho_m}\longrightarrow} \;
\cdots \; {\overset{-\rho_3}\longrightarrow}\; \pmb{X}_2 \;
{\overset{-\rho_2}\longrightarrow} \; \pmb{X}_1\; {\overset{-\rho_1}\longrightarrow} \; \pmb{X}_0 
\label{R.path}
\ee
in terms of the reversed sequence defined by also reversing the elementary reactions.  The probability of this reversed path is given by an expression similar to Eq.~(\ref{path-prob-2}). 


\section{Entropy production}
\label{Sec:entr-prod}

\subsection{Time evolution of entropy}

The expression of the entropy production rate can be obtained by considering the time evolution of entropy according to the stochastic process ruled by the chemical master equation~(\ref{master.eq}).  In this framework, the entropy can be defined as\cite{LVN84,G04,D15}
\be
S(t) = \sum_{\pmb{X}} \Big[ S^0(\pmb{X}) -  k_{\rm B}  \ln {\mathscr P}(\pmb{X},t) \Big] {\mathscr P}(\pmb{X},t) \, ,
\label{entropy}
\ee
where $S^0(\pmb{X})$ is the entropy of the system containing the molecular numbers $\pmb{X}$ for the intermediate species and $-k_{\rm B}\ln{\mathscr P}(\pmb{X},t)$ is the contribution to entropy due to the probability distribution over the different possibilities of molecular numbers $\pmb{X}$.  The time derivative of the total entropy~(\ref{entropy}) according to the reduced master equation~(\ref{master.eq}) can be decomposed as follows,\cite{LVN84,G04,D15}
\be
\frac{dS}{dt} = \frac{d_{\rm e}S}{dt}  +\frac{d_{\rm i}S}{dt} 
\label{entropy.varia}
\ee
into the flow of entropy exchange $d_{\rm e}S/dt$ between the system and its environment and the entropy production rate given by
\bea\label{EPR-stoch}
&&\frac{d_{\rm i}S}{dt} =k_{\rm B} \sum_{\pmb{X}}\sum_{\rho=1}^{r}  \Big[ W_{\rho}(\pmb{X}\vert\pmb{X}-\pmb{\nu}_{\rho})\, {\mathscr P}(\pmb{X}-\pmb{\nu}_{\rho},t)\nonumber\\
&&\qquad\qquad\qquad\qquad - W_{-\rho}(\pmb{X}-\pmb{\nu}_{\rho}\vert\pmb{X})\, {\mathscr P}(\pmb{X},t)\Big]  \nonumber\\
&&\qquad\qquad \times \ln \frac{W_{\rho}(\pmb{X}\vert\pmb{X}-\pmb{\nu}_{\rho})\, {\mathscr P}(\pmb{X}-\pmb{\nu}_{\rho},t)}{W_{-\rho}(\pmb{X}-\pmb{\nu}_{\rho}\vert\pmb{X})\, {\mathscr P}(\pmb{X},t)} \ge 0 \, . \qquad
\eea
This latter is always non negative in agreement with the second law of thermodynamics.  This entropy production rate evolves in time and is expected to reach its asymptotic value when the probability distribution converges towards the stationary solution ${\mathscr P}_{\rm st}(\pmb{X})$ of the reduced master equation~(\ref{master.eq}).  The entropy production rate is vanishing if the conditions of detailed balance are satisfied
\be
W_{\rho}(\pmb{X}\vert\pmb{X}-\pmb{\nu}_{\rho})\, {\mathscr P}_{\rm eq}(\pmb{X}-\pmb{\nu}_{\rho}) = W_{-\rho}(\pmb{X}-\pmb{\nu}_{\rho}\vert\pmb{X})\, {\mathscr P}_{\rm eq}(\pmb{X})\, ,
\ee
in which case the system is at thermodynamic equilibrium ${\mathscr P}_{\rm st}={\mathscr P}_{\rm eq}$.  Otherwise, the system is running in a {\it nonequilibrium steady state}.

\subsection{Random paths and entropy production}

 Recently, a method using the probabilities~(\ref{path-prob-2}) of the random paths~(\ref{path+n}) or~(\ref{path}) has been proposed to evaluate the entropy production rate in stochastic reactive systems.\cite{G04,AG04}  This method is in direct relation with Gillespie's algorithm\cite{G76,G77} simulating the coupled reactions of the network~(\ref{reactions}).  The method consists in comparing the probability of the random path~(\ref{path}) with the probability of its time reversal~(\ref{R.path}).  Because of Eq.~(\ref{path-prob-2}), the logarithm of the ratio of these probabilities is given by
\be
\Sigma(t) \equiv \ln \frac{{\mathsf P}_{\rm st}[{\cal X}(t)]}{{\mathsf P}_{\rm st}[{\cal X}^{\rm R}(t)]} = Z(t) + \ln \frac{{\mathscr P}_{\rm st}(\pmb{X}_0)}{{\mathscr P}_{\rm st}(\pmb{X}_m)}
\label{Sigma}
\ee
in terms of the quantity
\be
Z(t) \equiv \ln\prod_{k=1}^m \frac{W_{\rho_k}(\pmb{X}_k\vert\pmb{X}_{k-1})}{W_{-\rho_k}(\pmb{X}_{k-1}\vert\pmb{X}_k)} \, ,
\label{Z}
\ee
involving the transition rates of the forward and backward elementary reactions that have occurred during the path~(\ref{path}).  Indeed, the probabilities~(\ref{path-prob-2}) for the path and its time reversal contain the same exponential factors, which thus cancel each others, so that there remains the ratios of the rates of opposite transitions in Eq.~(\ref{Z}) and the ratio of the stationary probability distributions at both ends of the random path, giving the last term in Eq.~(\ref{Sigma}).  Under nonequilibrium conditions, the quantity~(\ref{Z}) is expected to grow linearly in time, while the last term of Eq.~(\ref{Sigma}) becomes negligible for long enough time.  Therefore, the mean growth rate of the quantity~(\ref{Sigma}) is determined by the quantity~(\ref{Z}):
\be\label{R}
R \equiv \lim_{t\to\infty} \frac{1}{t} \, \langle \Sigma(t)\rangle = \lim_{t\to\infty} \frac{1}{t} \, \langle Z(t)\rangle \, ,
\ee
where $\langle\cdot\rangle$ denotes the statistical average over the path probability distribution~(\ref{path-prob-2}).

Remarkably, the rate~(\ref{R}) is giving the entropy production rate~(\ref{EPR-stoch}) with the stationary distribution ${\mathscr P}_{\rm st}(\pmb{X})$:
\be\label{EPR-Z}
R = \frac{1}{k_{\rm B}}\, \frac{d_{\rm i}S}{dt}\bigg\vert_{\rm st} \ge 0 \, ,
\ee
as proved in Ref.~\onlinecite{G04}.  Accordingly, the entropy production can be evaluated using the quantity~(\ref{Z}) along the random paths~(\ref{path}) giving the sequence of elementary reactive events, which are simulated with Gillespie's algorithm.  This powerful method has been used in stochastic linear reaction networks,\cite{AG08PRE,TS08,XH10} Schl\"ogl's model for bistability,\cite{G04} the Brusselator model of chemical clocks,\cite{AG08JCP,XHX08} as well as other reaction networks.\cite{XHX09,RXH11,XZ18}  The method has also been tested for a stochastic linear reaction network, using diffusion approximations to the chemical master equation.\cite{H15}  Here below, this method will be applied to evaluate entropy production in the challenging case of chemical chaos.

Moreover, the quantity~(\ref{Z}) has been shown to obey a fluctuation theorem, from which Eq.~(\ref{EPR-Z}) can be deduced.\cite{G04,AG04,AG07}  These results have been extended in particular to the quantity~(\ref{Sigma}), leading to the recent development of stochastic thermodynamics.\cite{S05,XHX08,XHX09,XH10,RXH11,XZ18,S11,S12,D15,RE18a,RE18b}


\section{Deterministic limit}
\label{Sec:det-lim}

At macroscales, the molecular fluctuations become negligible in front of the macroscopic mean numbers of molecules of the different species.  In this limit, the macroscopic kinetic equations are recovered, as well as the expression of entropy production rate known in chemical nonequilibrium thermodynamics.\cite{D36,P67,N79,GM84,KP98}  Moreover, the Hill-Schnakenberg method of cycle decomposition can be used for the time average of the entropy production rate.\cite{S76,H05,BLG18} 

\subsection{Kinetic equations}

The macroscopic concentrations of the intermediate species can be defined as
\be
x_j \equiv \frac{1}{V} \, \langle X_j\rangle
\ee
with the mean numbers
\be\label{mean-x}
\langle X_j\rangle \equiv \sum_{\pmb{X}} X_j \, {\mathscr P}(\pmb{X},t) \, .
\ee
The reaction rates can be similarly defined as
\be\label{mean-w}
w_{\pm\rho} \equiv \frac{1}{V}\sum_{\pmb{X}} W_{\pm\rho}(\pmb{X}\pm\pmb{\nu}_{\rho}\vert\pmb{X}) \, {\mathscr P}(\pmb{X},t)
\ee
for $\rho=1,2,\dots,r$.  Equivalently, the reaction rates can be expressed as
\be
w_{\pm\rho} =\frac{1}{V}\, \frac{d}{dt}\, \langle n_{\pm\rho}(t) \rangle
\ee
in terms of the counters of the elementary chemical reactions and using the statistical average with respect to the joint probability distribution $P(\pmb{X},\pmb{n},t)$.

If the volume $V$ is large enough, the marginal distribution ${\mathscr P}(\pmb{X},t)$ may be supposed to be peaked around the most probable values for the molecular numbers.  In this limit, the reaction rates can be expressed as follows in terms of the most probable values for the concentrations $\pmb{x}=\{x_j\}_{j=1}^s$,
\be\label{macro_rates}
w_{\pm\rho}(\pmb{x})=  k_{\pm\rho}\; \prod_{i=1}^e a_i^{\tilde\nu_{i \rho}^{(\pm)} } \;  \prod_{j=1}^s x_j^{\nu_{j \rho}^{(\pm)}} 
\ee
for $\rho=1,2,\dots,r$, which are the known expressions for the macroscopic reaction rates according to the mass action law.\cite{GM84,N95}

As a consequence of Eq.~(\ref{DX}), the macroscopic kinetic equation for the concentration $x_j$ is given by
\be\label{kin.eq}
\frac{dx_j}{dt} = \sum_{\rho=1}^r \nu_{j \rho} \left(w_{+\rho}-w_{-\rho}\right)
\ee
in terms of the stoichiometric coefficients~(\ref{stoichio}).
Similarly, the mean rate of exchange with the exterior of the system in order to maintain fixed the concentration $a_i$ of species A$_i$ inside the volume $V$ can be deduced from Eq.~(\ref{DA}) giving
\be
\frac{d}{dt} \langle A_i\rangle = V \sum_{\rho=1}^r \tilde\nu_{i \rho} \left(w_{+\rho}-w_{-\rho}\right) .
\ee

\subsection{Macroscopic entropy production}

The probability distribution ${\mathscr P}(\pmb{X},t)$ being peaked around the most probable values $\pmb{X}=V\pmb{x}$ ruled by the kinetic equations~(\ref{kin.eq}), the entropy production rate~(\ref{EPR-stoch}) becomes
\be
\frac{d_{\rm i}S}{dt} = k_{\rm B} \, V \sum_{\rho=1}^{r} \left(w_{+\rho}-w_{-\rho}\right) \ln \frac{w_{+\rho}}{w_{-\rho}} \ge 0
\label{EPR-Ex}
\ee
in the macroscopic limit $V\to\infty$.  This is the expression expected from nonequilibrium chemical thermodynamics.\cite{D36,P67,N79,GM84,KP98} The sum is taken over all the elementary chemical reactions in the network.  The entropy production rate~(\ref{EPR-Ex}) is in general non negative, vanishing at the equilibrium concentrations $\pmb{x}_{\rm eq}$ such that the macroscopic conditions of detailed balance are satisfied: $w_{+\rho}(\pmb{x}_{\rm eq})=w_{-\rho}(\pmb{x}_{\rm eq})$.

\subsection{Time average and stoichiometric matrix}
\label{Subsec:Stoichio-matrix}

Defining the net reaction rates as
\be\label{w-rho}
w_{\rho} \equiv w_{+\rho}-w_{-\rho}
\ee
for $\rho=1,2,\dots,r$,  the coupled kinetic equations~(\ref{kin.eq}) can be written in the following form,
\be\label{det-dyn}
\frac{d\pmb{x}}{dt} = \pmb{\nu}\cdot\pmb{w}
\ee
using the vector $\pmb{w}=\{ w_{\rho}\}_{r=1}^r$ of reaction rates and the matrix of stoichiometric coefficients $\pmb{\nu}\equiv\{\pmb{\nu}_{\rho}\}_{\rho=1}^r = (\nu_{j\rho})_{j=1,\dots,s;\rho=1,\dots r}$.  

This set of first-order differential equations defines the {\it dynamical system} ruling the time evolution of the molecular concentrations in the phase space $\pmb{x}\in{\mathbb R}^{s}$.\cite{N95}  The solutions of these differential equations from given initial conditions define the flow of the dynamics in phase space: $\pmb{x}_t=\pmb{\Phi}^t\pmb{x}_0$.  In chemical kinetics, this dynamics is typically dissipative in the sense that the phase-space volumes are contracted on average, so that the trajectories are converging in time towards an attractor of zero Lebesgue measure in phase space.  This attractor may be a fixed point corresponding to a macroscopic stationary state, a limit cycle describing periodic oscillations, a torus if the oscillations are quasiperiodic, or a fractal attractor if the dynamics is chaotic and manifesting sensitivity to initial conditions characterized by positive Lyapunov exponents.\cite{N95}  

Even if the dynamics is time dependent, we can always consider the stationary probability density $p_{\rm st}(\pmb{x})$ associated with the attractor, using the time average
\be\label{time-aver}
\overline{(\cdot)} \equiv \lim_{T\to\infty} \frac{1}{T} \int_0^T (\cdot)\, dt \, .
\ee
Indeed, according to the Birkhoff-Khinchin ergodic theorem, this time average defines a stationary probability distribution, which almost always exists and has the trajectory issued from its initial condition as support.\cite{CFS82} Moreover, if the dynamics is unstable enough on the attractor, the property of ergodicity may hold and the stationary probability density can be constructed as a so-called SRB measure.\cite{ER85}  However, in the following, we shall only use the time average~(\ref{time-aver}).

A basic property of the time average~(\ref{time-aver}) is that
\be\label{Tav-dfdt}
\overline{\frac{df}{dt}}=0
\ee 
for the time derivative of any function $f=f(\pmb{x})$.  Therefore, we have that
\be\label{Tav-kin.eqs}
\overline{\frac{d\pmb{x}}{dt}} = \pmb{\nu}\cdot\overline{\pmb{w}}=0 \, .
\ee
Consequently, the time average of the reaction rates belongs to the right null space of the stoichiometric matrix.  This latter can be decomposed into the right null eigenvectors $\pmb{e}_c=\{e_{\rho c}\}_{\rho=1}^r$ such that
\be\label{RNE}
\pmb{\nu}\cdot\pmb{e}_c = 0 \qquad\mbox{for}\quad c=1,2,\dots,o \, .
\ee
Each right null eigenvector defines a cycle $c$, which is a sequence of elementary reactions (weighted by the components $e_{\rho c}$ of the eigenvector) such that the molecular numbers $\pmb{X}$ of intermediate species come back to their initial values once the cycle is closed.  We note that different sets of right null eigenvectors can be obtained using linear combinations.

Besides, the left null space of the stoichiometric matrix contains all the constants of motion $L=\pmb{\ell}^{\rm T}\cdot\pmb{x}$ such that $dL/dt=0$.  Accordingly, this space is spanned by the left null eigenvectors, $\pmb{\ell}^{\rm T}\cdot\pmb{\nu}=0$. If $l$ is the number of constants of motion, the rank of the stoichiometric matrix is given by\cite{PE14} 
\be
{\rm rank}(\pmb{\nu}) = s-l = r-o 
\ee
with the number $s$ of intermediate species, $r$ of reactions, and $o$ of cycles.

\subsection{Cycle decomposition}
\label{Subsec:Cycle-decomp}

Because of Eq.~(\ref{Tav-kin.eqs}), the time average of the reaction rates can be decomposed according to
\be
\overline{\pmb{w}} = \sum_{c=1}^o  \pmb{e}_c \, {\cal J}_c
\label{vec-w-decomp}
\ee
onto the basis of the $o$ right null eigenvectors~(\ref{RNE}).  The coefficients ${\cal J}_c$ of this decomposition define the rates associated with the $o$ cycles of the reaction network.  The components of Eq.~(\ref{vec-w-decomp}) are given by
\be
\overline{w}_{\rho} = \sum_{c=1}^o e_{\rho c} \, {\cal J}_c  \, .
\label{w-decomp}
\ee
The affinity associated with the $c^{\rm th}$ cycle is defined as
\be\label{Aff-cycle}
{\cal A}_c \equiv \ln \prod_{\rho=1}^r \left(\frac{w_{+\rho}}{w_{-\rho}}\right)^{e_{\rho c}} \, .
\ee
An important property is that these affinities do not depend on the concentrations $\pmb{x}$ of the intermediate species if the reaction rates obey the mass action law~(\ref{macro_rates}).  Indeed, the ratios of these opposite reaction rates can be written as follows in terms of the stoichiometric coefficients~(\ref{stoichio}):
\be
\frac{w_{+\rho}}{w_{-\rho}} = \frac{k_{+\rho}}{k_{-\rho}} \, \prod_{i=1}^e a_i^{-\tilde\nu_{i\rho}} \, \prod_{j=1}^s x_j^{-\nu_{j\rho}} \, .
\ee
Accordingly, the affinity~(\ref{Aff-cycle}) becomes
\bea
{\cal A}_c &=& \sum_{\rho=1}^r e_{\rho c} \, \ln \frac{w_{+\rho}}{w_{-\rho}} \\
&=& \sum_{\rho=1}^r e_{\rho c} \, \ln\left(\frac{k_{+\rho}}{k_{-\rho}} \, \prod_{i=1}^e a_i^{-\tilde\nu_{i\rho}}\right) -\sum_{j=1}^s\sum_{\rho=1}^r \nu_{j\rho}\, e_{\rho c} \, \ln x_j \, .\nonumber
\eea
Since $\sum_{\rho=1}^r \nu_{j\rho}\, e_{\rho c}=0$ because of Eq.~(\ref{RNE}), we find
\be\label{aff-reduced}
{\cal A}_c = \sum_{\rho=1}^r e_{\rho c} \, \ln\left(\frac{k_{+\rho}}{k_{-\rho}} \, \prod_{i=1}^e a_i^{-\tilde\nu_{i\rho}}\right) ,
\ee
thus establishing the property.

The remarkable result is that the time average of the macroscopic entropy production rate~(\ref{EPR-Ex}) can also be decomposed into the $o$ cycles if the reaction rates obey the mass action law~(\ref{macro_rates}).  For notational simplicity, we consider the dimensionless entropy production rate per unit volume given by
\be
\sigma\equiv \sum_{\rho=1}^{r} w_{\rho}\, \ln \frac{w_{+\rho}}{w_{-\rho}} \ge 0\, .
\label{sigma-det}
\ee
According to the mass action law, we have that
\be
\sigma= \sum_{\rho=1}^{r} w_{\rho}\, \ln\left(\frac{k_{+\rho}}{k_{-\rho}} \, \prod_{i=1}^e a_i^{-\tilde\nu_{i\rho}}\right) - \sum_{j=1}^s\sum_{\rho=1}^{r} \nu_{j\rho}\, w_{\rho}\, \ln x_j  \, .
\label{sigma-det-2}
\ee
Here, we note that the time derivative of the function
\be
f\equiv \sum_{j=1}^s x_j \ln x_j
\ee
is given by
\be
\frac{df}{dt} =  \sum_{j=1}^s \frac{dx_j}{dt} \left(\ln x_j + 1\right) =  \sum_{j=1}^s \sum_{\rho=1}^{r} \nu_{j\rho}\, w_{\rho} \left(\ln x_j + 1\right) ,
\ee
where we recognize the last term of Eq.~(\ref{sigma-det-2}).  According to Eqs.~(\ref{Tav-dfdt}) and~(\ref{Tav-kin.eqs}), we conclude that the time average of this last term is vanishing, so that the time average of Eq.~(\ref{sigma-det-2}) is given by
\be
\overline{\sigma}= \sum_{\rho=1}^{r} \overline{w}_{\rho}\, \ln\left(\frac{k_{+\rho}}{k_{-\rho}} \, \prod_{i=1}^e a_i^{-\tilde\nu_{i\rho}}\right) .
\label{av-sigma-det-1}
\ee
Now, using the decomposition~(\ref{w-decomp}), we obtain
\be
\overline{\sigma}= \sum_{c=1}^o {\cal J}_c \sum_{\rho=1}^{r} e_{\rho c} \, \ln\left(\frac{k_{+\rho}}{k_{-\rho}} \, \prod_{i=1}^e a_i^{-\tilde\nu_{i\rho}}\right)
\label{av-sigma-det-2}
\ee
Finally, using to the expression~(\ref{aff-reduced}) for the affinities, we find
\be
\overline{\sigma}= \sum_{c=1}^o {\cal A}_c \, {\cal J}_c  \, .
\label{av-sigma-det-3}
\ee
Hence, for reaction kinetics obeying the mass action law, the time average of the entropy production rate can always be decomposed as the sum of the affinities and rates associated with the $o$ cycles defined in terms of the stoichiometric matrix.  It should be pointed out that linear combinations may transform the sets of right null eigenvectors and, thus, the corresponding affinities and rates, but the entropy production rate~(\ref{av-sigma-det-3}) remains invariant under such transformations.

We note that all the affinities~(\ref{Aff-cycle}) are vanishing if the reaction network is at thermodynamic equilibrium.

\subsection{Entropy production by counting reactive events}

The result~(\ref{av-sigma-det-3}) is remarkable beyond the framework of the deterministic dynamics.  Indeed, in the stochastic approach, we may define the following random variable,
\be\label{Sigma-tilde}
\tilde\Sigma(t) \equiv \sum_{c=1}^o {\cal A}_c \, \Upsilon_c(t)
\ee
in terms of the affinities~(\ref{aff-reduced}) associated with the cycles and the corresponding signed cumulated numbers $\Upsilon_c(t)$.  These latter are related to the signed cumulated numbers $N_{\rho}\equiv n_{+\rho}-n_{-\rho}$ of elementary reactions [which have been introduced in Eq.~(\ref{DX})] according to
\be
N_{\rho}(t) = \sum_{c=1}^o e_{\rho c} \, \Upsilon_c(t)
\ee
in analogy with the decomposition~(\ref{w-decomp}).  Since the statistical averages of the signed cumulated numbers $\Upsilon_c(t)$ give the corresponding rates by
\be\label{Jc-Nc}
\tilde{\cal J}_c = \frac{1}{V} \, \lim_{t\to\infty} \frac{1}{t} \,  \langle\Upsilon_c(t)\rangle \, ,
\ee
we have that the entropy production rate can be evaluated as follows,
\be\label{Sigma-tilde-Jc}
\tilde{\sigma} = \frac{1}{V} \, \lim_{t\to\infty} \frac{1}{t} \, \langle\tilde\Sigma(t)\rangle = \sum_{c=1}^o {\cal A}_c \, \tilde{\cal J}_c \, ,
\ee
using the statistical average of the random variable~(\ref{Sigma-tilde}) defined for the stochastic process of Subsec.~\ref{Subsec:rndm-paths}.  This further result provides a complementary method to evaluate the entropy production rate directly by counting the numbers of reactive events occurring along the paths~(\ref{path+n}) simulated by Gillespie's algorithm, besides the method using mean growth rate of the quantity~(\ref{Z}).  We note that the random variable~(\ref{Sigma-tilde}) obeys a fluctuation theorem as a consequence of the multivariate fluctuation theorem for currents proved in Ref.~\onlinecite{AG07}.


\section{Model of chemical chaos}
\label{Sec:GN83}

\subsection{The reaction network}

We consider the reversible version of the model of chemical chaos proposed in Ref.~\onlinecite{GN83}.  This model has $s=3$ intermediate species X~=~X$_1$, Y~=~X$_2$, and Z~=~X$_3$, which is the minimum number required in order to have chaotic behavior in the deterministic dynamics.  The reaction network of the model is defined according to
\bea
{\rm A}_1 + {\rm X}\, &\underset{k_{-1}}{\overset{k_{+1}}{\rightleftharpoons}}&\, 2\, {\rm X} \, ,\nonumber \\
{\rm A}_2 + 2\, {\rm X}\, &\underset{k_{-2}}{\overset{k_{+2}}{\rightleftharpoons}}&\, 3\, {\rm X} \, ,\nonumber \\
{\rm X} + {\rm Y}\, &\underset{k_{-3}}{\overset{k_{+3}}{\rightleftharpoons}}&\, {\rm A}_3 + {\rm Y}\, , \nonumber \\
{\rm X} + {\rm Z}\, &\underset{k_{-4}}{\overset{k_{+4}}{\rightleftharpoons}}&\, {\rm A}_4 + {\rm Z}\, , \nonumber \\
{\rm A}_5 + {\rm X} +  {\rm Y}\, &\underset{k_{-5}}{\overset{k_{+5}}{\rightleftharpoons}}&\, {\rm X}+2\, {\rm Y} \,  , \nonumber\\
{\rm A}_6 + {\rm Y} +  {\rm Z}\, &\underset{k_{-5}}{\overset{k_{+5}}{\rightleftharpoons}}&\, 2\, {\rm Y} + {\rm Z} \,  , \nonumber\\
{\rm Y}\, &\underset{k_{-7}}{\overset{k_{+7}}{\rightleftharpoons}}&\, {\rm A}_7 \, , \nonumber \\
{\rm A}_8 + {\rm X}\, &\underset{k_{-8}}{\overset{k_{+8}}{\rightleftharpoons}}&\, {\rm X} + {\rm Z} \, ,\nonumber \\
{\rm A}_9 + 2\, {\rm Z}\, &\underset{k_{-9}}{\overset{k_{+9}}{\rightleftharpoons}}&\, 3 \, {\rm Z}\, , \nonumber\\
{\rm Z}\, &\underset{k_{-10}}{\overset{k_{+10}}{\rightleftharpoons}}&\, {\rm A}_{10} \, . \label{GN83-CRN}
\eea
This model forms a network of $r=10$ elementary forward reactions and so many backward reactions.  Several elementary reactions are trimolecular.  The kinetics is supposed to obey the mass action law with the reaction constants $k_{\pm\rho}$ for $\rho=1,2,\dots,r$.  Every one of the ten reactions involves a reactant or production species A$_i$ ($i=\rho=1,2,\dots,r$), which is chemostatted.  

\subsection{Macroscopic kinetics and chaos}

Their macroscopic reaction rates are given by
\bea
&& w_{+1} = k_{+1}\,  a_{+1}\,  x \, , \qquad\   w_{-1}=k_{-1}\,  x^2 \, , \label{w1} \\
&& w_{+2} = k_{+2}\,  a_{+2}\,  x^2 \, , \qquad w_{-2}=k_{-2}\,  x^3 \, , \label{w2} \\
&& w_{+3} = k_{+3}\,  x\,  y \, , \qquad\quad\ w_{-3}=k_{-3}\,  a_3\,  y \, , \label{w3} \\
&& w_{+4} = k_{+4}\,  x\,  z \, , \qquad\quad\; w_{-4}=k_{-4}\,  a_4\,  z \, , \label{w4} \\
&& w_{+5} = k_{+5}\,  a_5\,  x\,  y \, , \qquad\, w_{-5}=k_{-5}\,  x\,  y^2 \, , \label{w5} \\
&& w_{+6} = k_{+6}\,  a_6\,  y\,  z \, , \qquad\, w_{-6}=k_{-6}\,  y^2\,  z \, , \label{w6} \\
&& w_{+7} = k_{+7}\,  y \, , \qquad\qquad w_{-7}=k_{-7}\,  a_7 \, , \label{w7} \\
&& w_{+8} = k_{+8}\,  a_8\,  x \, , \qquad\ \ w_{-8}=k_{-8}\,  x\,  z \, , \label{w8} \\
&& w_{+9} = k_{+9}\,  a_9\,  z^2 \, , \qquad\ w_{-9}=k_{-9}\,  z^3 \, , \label{w9} \\
&& w_{+10} = k_{+10}\,  z \, , \qquad\quad w_{-10}=k_{-10}\, a_{10} \, , \label{w10}
\eea
in terms of the concentrations $(x,y,z)$ of the intermediate species $({\rm X},{\rm Y},{\rm Z})$.
According to Eq.~(\ref{kin.eq}) with the notation~(\ref{w-rho}), the kinetic equations for the concentrations of the intermediate species take the following form,
\bea
\dot{x} &=& w_1 + w_2 -w_3 -w_4 \, , \label{dot-x}\\
\dot{y} &=& w_5 + w_6 -w_7 \, , \label{dot-y}\\
\dot{z} &=& w_8 + w_9 -w_{10} \, . \label{dot-z}
\eea

These differential equations can be numerically integrated using a Runge-Kutta algorithm of orders 4 and 5 with a variable time step.  The following parameter values are chosen:
\bea
&& k_{+1}=0.6 \, , \ 0.3<k_{+2} < 0.7 \, , k_{+3}=0.5 \, , k_{+4}=k_{+5}=1 \, , \nonumber\\
&& k_{+6}=0.3 \, , k_{+7}=1.3 \, ,  k_{+8}=100 \, , k_{+9}=300 \, , k_{+10}=500  \, ,  \nonumber\\
&& k_{-9}=50 \, , \  k_{-\rho}=0.01 \ \ \mbox{for} \ \ \rho=1\mbox{-}8,10 \, , \nonumber\\
&&\mbox{and} \quad a_i=1 \ \ \mbox{for} \ \ i=1\mbox{-}10 \, . \label{parameters}
\eea

\begin{figure}[h]
\centerline{\scalebox{0.5}{\includegraphics{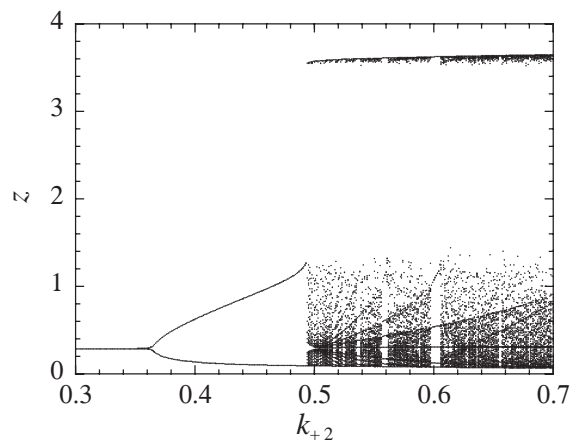}}}
\caption{Bifurcation diagram of the dynamical system of Eqs.~(\ref{dot-x})-(\ref{dot-z}) with the reaction rates~(\ref{w1})-(\ref{w10}) and the parameter values~(\ref{parameters}).  The axis $k_{+2}$ is scanned every step $\Delta k_{+2}= 10^{-3}$.  The dots depict every extremum where $\dot{z}(t)=0$ in the time interval $500<t<600$, after transients over the time interval $0<t<500$.}
\label{fig1}
\end{figure}

\begin{figure*}[ht]
\centerline{\scalebox{0.47}{\includegraphics{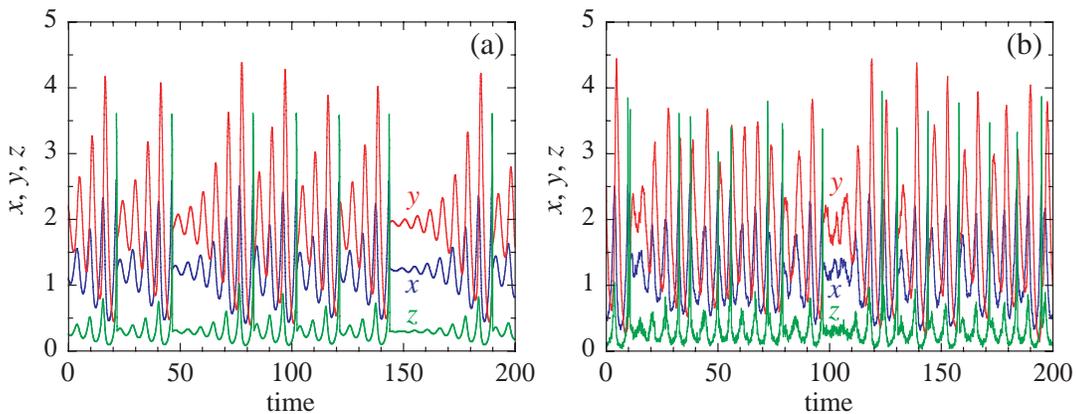}}}
\caption{(a) The concentrations $(x,y,z)$ versus time $t$ for the dynamical system of Eqs.~(\ref{dot-x})-(\ref{dot-z}) with the reaction rates~(\ref{w1})-(\ref{w10}) and the parameter values~(\ref{parameters}) with $k_{+2}=0.55$.  (b) The concentrations $(x,y,z)$ versus time $t$ for the corresponding stochastic process with the extensivity parameter $V=250$.}
\label{fig2}
\end{figure*}

\begin{figure*}[ht]
\centerline{\scalebox{0.47}{\includegraphics{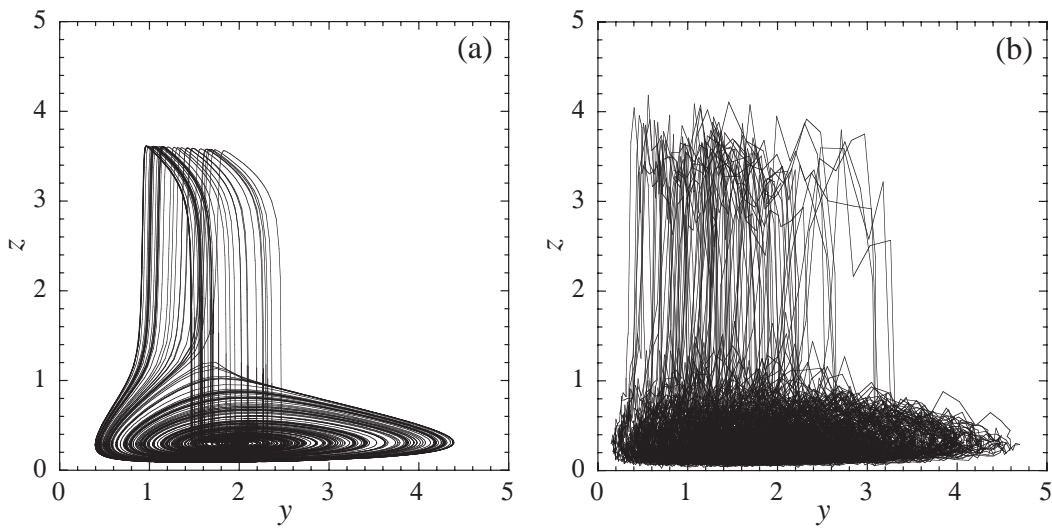}}}
\caption{(a) The chaotic attractor formed by the trajectory of Fig.~\ref{fig2}(a) for $100<t<1100$. (b) The corresponding noisy attractor formed by the trajectory of Fig.~\ref{fig2}(b) for $0<t<1000$ with the extensivity parameter $V=250$.}
\label{fig3}
\end{figure*}

The corresponding bifurcation diagram is shown in Fig.~\ref{fig1}.  We observe the following regimes:
\bea
&&\mbox{stationary:} \qquad\qquad\quad\ \, k_{+2} \lesssim 0.363 \, , \nonumber\\
&&\mbox{periodic:} \qquad\ \ \, 0.363 \lesssim k_{+2} \lesssim 0.494 \, , \nonumber\\
&&\mbox{chaotic:} \qquad\quad\; 0.494 \lesssim k_{+2}  \, , \nonumber
\eea
with windows of mixed-mode oscillations in the chaotic regime.

For the value $k_{+2}=0.55$, a trajectory and the attractor are plotted in Fig.~\ref{fig2}(a) and Fig.~\ref{fig3}(a), respectively.  We see that the dynamics is chaotic for this parameter value.  The attractor in Fig.~\ref{fig3}(a) shows that chaos arises due to R\"ossler's mechanism of reinjection in phase space.  For the parameter values~(\ref{parameters}), the equation $\dot{z}=0$ defines a sigmoidal slow manifold in phase space, allowing the reinjection in the vicinity of the saddle focus on the lower branch of the slow manifold.  The trajectories are spiraling out of this saddle focus.  Once the amplitude of the oscillations is large enough, the trajectories are jumping to the upper branch of the slow manifold, where they move towards its edge, before being reinjected back onto the lower branch.  This reinjection allows the existence of a homoclinic orbit of Shil'nikov type, which is known to generate unstable periodic orbits of arbitrarily long period, as well as Smale's horseshoes of chaotic subsets.\cite{S65}

\subsection{Cycle decomposition}

As explained in Subsecs.~\ref{Subsec:Stoichio-matrix} and~\ref{Subsec:Cycle-decomp}, the time averages of the reaction rates and entropy production rate can be decomposed into the cycles given by the right null eigenvectors of the stoichiometric matrix.  For the reaction network~(\ref{GN83-CRN}), this latter is given by
\bea
&&\qquad w_1 \ w_2 \ \; w_3 \ \ w_4 \ \, w_5 \ w_6 \ \; w_7 \ w_8 \ w_9 \ w_{10} \nonumber \\
\pmb{\nu} &=& 
\begin{array}{r}
x \\
y \\
z 
\end{array}
\left(
\begin{array}{rrrrrrrrrr}
1 &\ 1 & -1 & -1 & \ 0 &\ 0 & 0 &\ 0 &\ 0 & 0 \\
0 &\ 0 & 0 & 0 & \ 1 &\ 1 & -1 &\ 0 &\ 0 & 0 \\
0 &\ 0 & 0 & 0 & \ 0 &\ 0 & 0 &\ 1 &\ 1 & -1 
\end{array}
\right) . \qquad
\eea
Looking for its left and right null eigenvectors, we find that the reaction network has $l=0$ constant of motion and $o=7$ cycles, so that the rank of the stoichiometric matrix is equal to
\be
{\rm rank}(\pmb{\nu}) = s-l = r-o = 3 \, .
\ee
The right null eigenvectors~(\ref{RNE}) associated with the $o=7$ cycles are listed here below together with their affinities~(\ref{Aff-cycle}), their rates given by Eq.~(\ref{vec-w-decomp}), and their corresponding overall reaction:
\bea
\mbox{cycle 1}: \quad && \pmb{e}_1^{\rm T} = (1,-1,0,0,0,0,0,0,0,0) \, , \nonumber\\
&& {\cal A}_1 = \ln \frac{w_{+1}w_{-2}}{w_{-1}w_{+2}} = \ln \frac{k_{+1}a_1k_{-2}}{k_{-1}k_{+2}a_2} \, , \nonumber\\
&& {\cal J}_1 =  -\overline{w}_2  \, , \qquad\qquad {\rm A}_1 \to {\rm A}_2 \, ;  \label{cycle1}
\eea
\bea
\mbox{cycle 2}: \quad && \pmb{e}_2^{\rm T} = (1,0,1,0,0,0,0,0,0,0) \, , \nonumber\\
&& {\cal A}_2 = \ln \frac{w_{+1}w_{+3}}{w_{-1}w_{-3}} = \ln \frac{k_{+1}a_1k_{+3}}{k_{-1}k_{-3}a_3} \, , \nonumber\\
&& {\cal J}_2 =  \overline{w}_3  \, , \qquad\qquad {\rm A}_1 \to {\rm A}_3 \, ;  \label{cycle2}
\eea
\bea
\mbox{cycle 3}: \quad && \pmb{e}_3^{\rm T} = (1,0,0,1,0,0,0,0,0,0) \, , \nonumber\\
&& {\cal A}_3 = \ln \frac{w_{+1}w_{+4}}{w_{-1}w_{-4}} = \ln \frac{k_{+1}a_1k_{+4}}{k_{-1}k_{-4}a_4} \, , \nonumber\\
&& {\cal J}_3 =  \overline{w}_4  \, , \qquad\qquad {\rm A}_1 \to {\rm A}_4 \, ;  \label{cycle3}
\eea
\bea
\mbox{cycle 4}: \quad && \pmb{e}_4^{\rm T} = (0,0,0,0,1,-1,0,0,0,0) \, , \nonumber\\
&& {\cal A}_4 = \ln \frac{w_{+5}w_{-6}}{w_{-5}w_{+6}} = \ln \frac{k_{+5}a_5 k_{-6}}{k_{-5}k_{+6}a_6} \, , \nonumber\\
&& {\cal J}_4 =  -\overline{w}_6  \, , \qquad\qquad {\rm A}_5 \to {\rm A}_6  \, ;  \label{cycle4}
\eea
\bea
\mbox{cycle 5}: \quad && \pmb{e}_5^{\rm T} = (0,0,0,0,1,0,1,0,0,0) \, , \nonumber\\
&& {\cal A}_5 = \ln \frac{w_{+5}w_{+7}}{w_{-5}w_{-7}} = \ln \frac{k_{+5}a_5 k_{+7}}{k_{-5}k_{-7}a_7} \, , \nonumber\\
&& {\cal J}_5 =  \overline{w}_7 \, , \qquad\qquad {\rm A}_5 \to {\rm A}_7 \, ;  \label{cycle5}
\eea
\bea
\mbox{cycle 6}: \quad && \pmb{e}_6^{\rm T} = (0,0,0,0,0,0,0,1,-1,0) \, , \nonumber\\
&& {\cal A}_6 = \ln \frac{w_{+8}w_{-9}}{w_{-8}w_{+9}} = \ln \frac{k_{+8}a_8 k_{-9}}{k_{-8}k_{+9}a_9} \, , \nonumber\\
&& {\cal J}_6 =  -\overline{w}_9 \, , \qquad\qquad {\rm A}_8 \to {\rm A}_9 \, ;  \label{cycle6}
\eea
\bea
\mbox{cycle 7}: \quad && \pmb{e}_7^{\rm T} = (0,0,0,0,0,0,0,1,0,1) \, , \nonumber\\
&& {\cal A}_7 = \ln \frac{w_{+8}w_{+10}}{w_{-8}w_{-10}} = \ln \frac{k_{+8}a_8 k_{+10}}{k_{-8}k_{-10}a_{10}} \, , \nonumber\\
&& {\cal J}_7 =  \overline{w}_{10} \, , \qquad\qquad {\rm A}_8 \to {\rm A}_{10} \, . \label{cycle7}
\eea
The explicit calculation of the affinities~(\ref{Aff-cycle}) confirms that they do not depend on the concentrations $(x,y,z)$ of the intermediate species, as expected according to Eq.~(\ref{aff-reduced}).  Moreover, Eq.~(\ref{vec-w-decomp}) allows us to relate the time averages of the reaction rates to the rates ${\cal J}_c$ associated with the cycles.  In addition to the reaction rates appearing in Eqs.~(\ref{cycle1})-(\ref{cycle7}), we also have the following relations for the other rates:  $\overline{w}_1={\cal J}_1+{\cal J}_2+{\cal J}_3$, $\overline{w}_5={\cal J}_4+{\cal J}_5$, and $\overline{w}_8={\cal J}_6+{\cal J}_7$, whereupon the time averages of the kinetic equations~(\ref{dot-x})-(\ref{dot-z}) are equal to zero, as required.

As a consequence of Eq.~(\ref{av-sigma-det-3}), the time average of the entropy production rate is obtained as follows,
\bea\label{EPR-det-GN83}
\overline{\sigma} &=& -{\cal A}_1 \overline{w}_2 + {\cal A}_2 \overline{w}_3 + {\cal A}_3 \overline{w}_4 \nonumber\\ && - {\cal A}_4 \overline{w}_6 + {\cal A}_5 \overline{w}_7 - {\cal A}_6 \overline{w}_9 + {\cal A}_7 \overline{w}_{10} \, .
\eea

\subsection{Simulating the stochastic process}

For the reaction network~(\ref{GN83-CRN}), the stochastic process can be defined in terms of the transition rates deduced with Eq.~(\ref{rates_mass_action}) and corresponding to the macroscopic rates~(\ref{w1})-(\ref{w10}).  This Markovian process can be simulated using Gillespie's algorithm.\cite{G76,G77} At every transition, the random variables are updated according to
\bea
&& t'=t+ \Delta t \, \nonumber\\
&& \pmb{X'}=\pmb{X}+ \pmb{\nu}_{\rho} \, , \nonumber\\
&& \pmb{n'} = \pmb{n} + \Delta\pmb{n}_{\rho} \, ,  \label{transition-Gillespie}
\eea
where the waiting time $\Delta t$ before the transition has the following exponential probability distribution of parameter given by the escape rate~(\ref{esc_rate}),
\be
p(\Delta t) = \gamma(\pmb{X}) \, {\rm e}^{-\gamma(\pmb{X})\, \Delta t} \, ,
\ee
$\pmb{\nu}_{\rho}=\{\nu_{j\rho}\}_{j=1}^s$ are the stoichiometric coefficients of the intermediate species, and $\Delta\pmb{n}_{\rho}=\{\delta_{\rho\rho'}\}_{\rho'=\pm 1}^{\pm r}$ are the jumps of the reaction counters upon the reactive event $\rho$.  This stochastic process is ruled by the master equation~(\ref{full.master.eq}).  Moreover, the time evolution of the quantity~(\ref{Z}) is simulated during the process with the following addition at every transition~(\ref{transition-Gillespie}):
\be\label{Z-jump}
Z'=Z+ \ln \frac{W_{\rho}(\pmb{X'}\vert\pmb{X})}{W_{-\rho}(\pmb{X}\vert\pmb{X'})}  \, .
\ee

A stochastic trajectory for the concentrations $\pmb{x}=\pmb{X}/V=(x,y,z)$ of the intermediate species simulated with the Gillespie algorithm is shown in Fig.~\ref{fig2}(b) for the value $V=250$ of the extensivity parameter entering the expressions~(\ref{rates_mass_action}) of the transition rates.  We observe that the trajectory is noisy, but keep the main features of the deterministic trajectory depicted in Fig.~\ref{fig2}(a).  In particular, the noisy oscillations have amplitude ranging between similar values and the episodic oscillations of small amplitude are also manifesting themselves.  The stochastic trajectory is plotted in Fig.~\ref{fig3}(b) in the same plane $(y,z)$ as in Fig.~\ref{fig3}(a), depicting a noisy attractor.  Here also, the key features of the deterministic attractors are observed, namely noisy oscillations close to the lower branch of the slow manifold with random jumps towards the upper branch, and random reinjections back onto the lower one.  Noise tends to disperse the excursions of the stochastic trajectory beyond the region of the attractor, but this dispersion is mild, as seen in Fig.~\ref{fig3}(b).  Accordingly, the features of chemical chaos are robust under the effect of stochasticity in this model.

\subsection{Entropy production rate}

Here, our aim is to compare the different methods to evaluate the entropy production rate in the deterministic and stochastic approaches.  The evaluation is performed in units with Boltzmann's constant is equal to unity: $k_{\rm B}=1$.

For the deterministic system, the time average of the entropy production rate~(\ref{sigma-det}) is obtained by integrating the ordinary differential equations~(\ref{dot-x})-(\ref{dot-z}) using a Runge-Kutta algorithm of orders 4 and 5 with a variable time step.  Furthermore, the time averages of the rates ${\cal J}_c$ associated with the seven cycles~(\ref{cycle1})-(\ref{cycle7}) can also be calculated to get the value~(\ref{av-sigma-det-3}).  The comparison between these values shows agreement within great numerical accuracy. In particular, for $k_{+2}=0.55$ in the chaotic regime, we find $\overline{\sigma}=3116.06$ and $\sum_{c=1}^7{\cal A}_c{\cal J}_c=3116.05$ after averaging over the time interval $t=1000$.  Therefore, these numerical calculations confirm the validity of the result~(\ref{av-sigma-det-3}), according to which the time average of the entropy production rate can be decomposed into the contributions of the cycles for the deterministic chaotic dynamics.

\begin{table}
\caption{\label{table} Contributions to entropy production of the reaction network~(\ref{GN83-CRN}) for the parameter values~(\ref{parameters}) with $k_{+2}=0.55$.  For the deterministic system, the values are computed by integrating over the time interval $t=1000$ with a Runge-Kutta algorithm of orders 4 and 5 with a variable time step.  The time average of the entropy production rate~(\ref{sigma-det}) for the deterministic system takes the rounded value $\overline{\sigma}=3116$ and the same rounded value is given by Eq.~(\ref{av-sigma-det-3}).  For the stochastic process, Gillespie's algorithm is used over a time interval $t=1000$ for $V=100$, $150$, $200$, and $250$ and the given values are obtained by extrapolation.  The mean growth rate~(\ref{R}) and~Eq.~(\ref{Sigma-tilde-Jc}) give the value $R/V=\tilde\sigma=3177\pm 180$.  In the table, ${\cal A}_c$ denotes the affinity~(\ref{aff-reduced}) associated with the cycle~$c$, ${\cal J}_c$ the time average of the corresponding deterministic rate given in Eqs.~(\ref{cycle1})-(\ref{cycle7}), $\tilde{\cal J}_c$ the mean value of the stochastic rate obtained with Eq.~(\ref{Jc-Nc}).  The confidence interval is estimated from the values obtained for each quantity in the simulations with $V=100$, $150$, $200$, and $250$.}
\vspace{5mm}
\begin{center}
\begin{tabular}{|ccccc|}
\hline
$c$  & ${\cal A}_c$ & ${\cal J}_c$ & ${\cal A}_c{\cal J}_c$ & ${\cal A}_c\tilde{\cal J}_c$ \\
\hline
$1$ & $0.087$ & $-0.904$ & $-0.0786$  & $-0.079\pm 0.003$  \\
$2$ & $8.006$  & $1.105$ & $8.85$ & $8.6\pm 0.4$  \\
$3$ & $8.70$ & $0.512$ & $4.45$ & $4.7\pm 0.6$  \\
$4$ & $1.20$ & $-0.173$ & $-0.208$ & $-0.21\pm 0.01$  \\
$5$ & $9.47$ & $2.37$ & $22.4$ & $21.9\pm 1.0$  \\
$6$ & $7.42$ & $-50.8$ & $-376.9$ & $-423\pm 110$  \\
$7$ & $20.03$ & $172.6$ & $3457.2$ & $3565\pm 290$  \\
\hline
$\sum_{c=1}^7{\cal A}_c{\cal J}_c$ & & &$3116$ & $3177\pm 180$  \\
\hline
\end{tabular}
\end{center}
\end{table} 

In the stochastic approach, the entropy production rate can be evaluated as the mean growth rate of the quantity~(\ref{Z}), which is given in Gillespie's algorithm with Eq.~(\ref{Z-jump}), or as the mean growth rate of the sum~(\ref{Sigma-tilde}) of the affinities $\{{\cal A}_c\}_{c=1}^7$ and the signed cumulated numbers $\{\Upsilon_c(t)\}_{c=1}^7$ associated with the cycles.  According to Eqs.~(\ref{cycle1})-(\ref{cycle7}), these numbers are given in terms of the signed cumulated numbers of elementary reactions $\{N_{\rho}=n_{+\rho}-n_{-\rho}\}_{\rho=1}^{10}$ respectively by $\Upsilon_1=-N_2$, $\Upsilon_2=N_3$, $\Upsilon_3=N_4$, $\Upsilon_4=-N_6$, $\Upsilon_5=N_7$, $\Upsilon_6=-N_9$, and $\Upsilon_7=N_{10}$.

These mean growth rates are computed using Gillespie's algorithm.  First, Eqs.~(\ref{R}) and~(\ref{Sigma-tilde-Jc}) give equal values $R=\tilde\sigma V$ for the entropy production rate up to great numerical accuracy, which confirms the validity of the decomposition into cycles for the entropy production rate in the stochastic approach.  Next, in order to compare with the deterministic value of the entropy production rate, the mean growth rates are calculated for increasing values of the extensivity parameter: $V=100$, $150$, $200$, and $250$.  The so-computed rates are used for extrapolating the value in the limit $V\to\infty$ by supposing a convergence as $V^{-1}$.  For $k_{+2}=0.55$ in the chaotic regime shown in Fig.~\ref{fig2}(b) and Fig.~\ref{fig3}(b),  the different rates and corresponding contributions to the entropy production rate are given in Table~\ref{table}.  

\begin{figure}[h]
\centerline{\scalebox{0.5}{\includegraphics{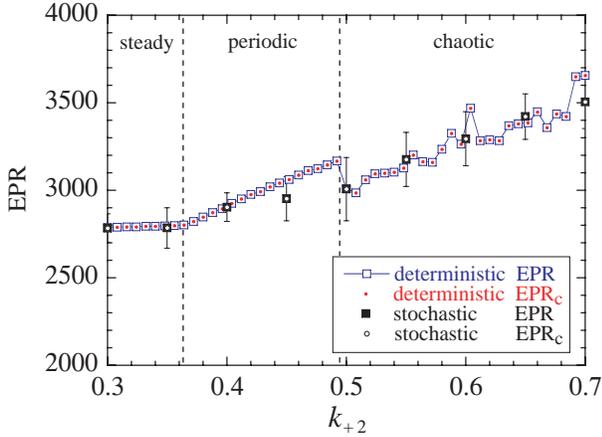}}}
\caption{Entropy production rate (EPR) of the reaction network~(\ref{GN83-CRN}) for the parameter values~(\ref{parameters}) versus $k_{+2}$, as computed with different methods in the deterministic and stochastic approaches.  The deterministic EPR (open squares) is obtained with the time average of the entropy production rate~(\ref{sigma-det}).  The deterministic EPR$_{\rm c}$ (filled circles) is calculated by the sum~(\ref{av-sigma-det-3}) over the $o=7$ cycles of the reaction network.  The stochastic EPR (filled squares) is computed by extrapolation of the values $R/V$ given by Eq.~(\ref{R}) using Gillespie's algorithm with $V=100$, $150$, $200$, and $250$. The stochastic EPR$_{\rm c}$ (open circles) is similarly obtained from Eq.~(\ref{Sigma-tilde-Jc}) in the same simulations using Gillespie's algorithm.  The averages are carried out over the time interval $t=1000$.}
\label{fig4}
\end{figure}

Furthermore, the deterministic and stochastic entropy production rates (EPR) are computed as a function of the parameter $k_{+2}$, as shown in Fig.~\ref{fig4}.  For the deterministic dynamics, we see that the time average of the entropy production rate obtained with Eq.~(\ref{sigma-det}) and Eq.~(\ref{av-sigma-det-3}) coincide within great numerical accuracy.  The dependence of the entropy production rate versus $k_{+2}$ is smooth in the stationary and periodic regimes.  However, we observe in Fig.~\ref{fig4} an irregular dependence in the chaotic regime.  The reason is that the attractor is not structurally stable in this regime, but it undergoes complex bifurcations between chaotic and periodic attractors, as seen in Fig.~\ref{fig1}.  The orbits include oscillations of small and large amplitudes, which is a characteristic feature of mixed-mode oscillations.  Accordingly, the time average of the entropy production rate undergoes variations due to the bifurcations between these chaotic and periodic attractors.  The general trend is the increase of the entropy production rate with the parameter $k_{+2}$.  In the stochastic approach, the mean entropy production rate is evaluated from the mean growth rate~(\ref{R}) of the quantity~(\ref{Z}) (divided by the extensivity parameter $V$), as well as from Eq.~(\ref{Sigma-tilde-Jc}) based on the cycle decomposition.  Again, both methods agree with each other up to great numerical accuracy.  In order to compare with the deterministic results, the mean values of the entropy production rate are calculated by extrapolations from the values obtained in simulations with Gillespie's algorithm for $V=100$, $150$, $200$, and $250$.  The values obtained with the stochastic approach are plotted in Fig.~\ref{fig4}, showing agreement with the deterministic values up to the statistical errors of the stochastic simulations.

Therefore, the rate of entropy production can be calculated consistently in both the deterministic and stochastic approaches.  The essential aspect is the role of elementary chemical reactions, as emphasized with the method based on the cycle decomposition where the entropy production rate is directly obtained by counting the reactive events due to elementary chemical reactions.

\subsection{Changing the direction of chemical conversion}

In the chaotic regime for $k_{+2}=0.55$, Table~\ref{table} shows that the affinity and the mean rate are opposite to each other for the cycles $c=1$, $c=4$, and $c=6$, so that these cycles have negative contributions to the entropy production rate.  This is consistent with the second law of thermodynamics because the other cycles have large enough positive contributions, so that the sum of all the contributions gives a positive entropy production, as required.  Thus, the cycles $c=1$, $c=4$, and $c=6$ are running in the direction that is opposite to the one of their spontaneous relaxation towards equilibrium.  These particular cycles of reactions are driven in the opposite direction because of their coupling to the other cycles in the network.

Now, we note that the parameter $k_{+2}$ only appears in the affinity ${\cal A}_1$ of the first cycle~(\ref{cycle1}).  Therefore, if $k_{+2}$ is tuned as a control parameter, the affinity ${\cal A}_1$ is changing, although the values of all the other affinities remain constant.  For the set~(\ref{parameters}) of parameter values, this affinity is given by ${\cal A}_1=\ln(0.6/k_{+2})$, so it is positive if $k_{+2}<0.6$ to become negative if $k_{+2}>0.6$.  Since the mean rate ${\cal J}_1$ remains negative on both sides of $k_{+2}=0.6$, the contribution ${\cal A}_1{\cal J}_1$ to the entropy production is changing sign if the control parameter crosses the value $k_{+2}=0.6$.  The cycle is running in the direction of the conversion ${\rm A}_1\to{\rm A}_2$ for $k_{+2}>0.6$, but in the direction of the opposite conversion ${\rm A}_2\to{\rm A}_1$ for $k_{+2}<0.6$, while remaining in the chaotic regime.  Remarkably, the direction of this chemical conversion can be controlled by tuning the parameter $k_{+2}$.  

\begin{figure}[b]
\centerline{\scalebox{0.5}{\includegraphics{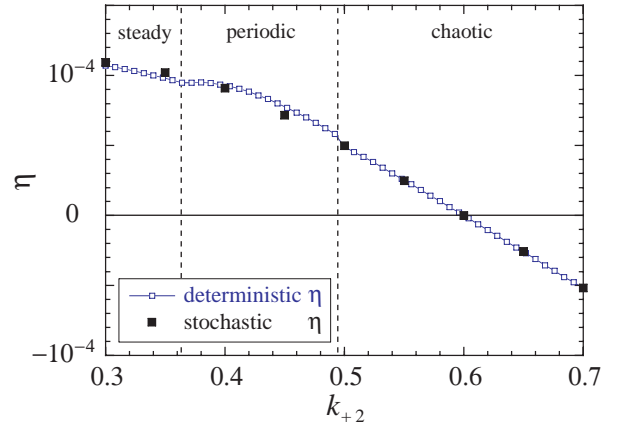}}}
\caption{Efficiency~(\ref{eta}) of the opposite chemical conversion ${\rm A}_2\to{\rm A}_1$ in the reaction network~(\ref{GN83-CRN}) for the parameter values~(\ref{parameters}) versus $k_{+2}$, as computed in the deterministic and stochastic approaches.  The deterministic values of the efficiency $\eta$ (open squares) are obtained using the time averages of the cycles~(\ref{cycle1})-(\ref{cycle7}).  The stochastic values of $\eta$ (filled squares) are computed by extrapolating the values obtained using Gillespie's algorithm with $V=100$, $150$, $200$, and $250$. The averages are carried out over the time interval $t=1000$.  The statistical errors on the stochastic values are here smaller than or the same size as the filled squared.}
\label{fig5}
\end{figure}

Since the reactive system is out of equilibrium, the chemical conversion ${\rm A}_2\to{\rm A}_1$, which has a direction opposite to the spontaneous one, is driven by the other reactions in the system.  Like in engines, the efficiency of this opposite conversion can be introduced as
\be\label{eta}
\eta\equiv - \frac{{\cal A}_1{\cal J}_1}{\sum_{c=2}^7 {\cal A}_c{\cal J}_c} \, .
\ee
This quantity is positive in the regime of the conversion ${\rm A}_2\to{\rm A}_1$, where it characterizes its efficiency.  According to the second law of thermodynamics requiring that the entropy production rate~(\ref{av-sigma-det-3}) or~(\ref{Sigma-tilde-Jc}) is always non negative, the efficiency~(\ref{eta}) is bounded as $\eta\le 1$.  In the stochastic approach, the efficiency is given by Eq.~(\ref{eta}) with the rates~$\tilde{\cal J}_c$ defined by Eq.~(\ref{Jc-Nc}).

The efficiency~(\ref{eta}) can be computed in the deterministic and stochastic approaches, as shown in Fig.~\ref{fig5}.  We see that the values obtained in both approaches are in agreement.  The efficiency is indeed observed to change sign as expected at the parameter value $k_{+2}=0.6$, confirming that the chemical conversion proceeds in the opposite direction ${\rm A}_2\to{\rm A}_1$ for $k_{+2}<0.6$, but in spontaneous direction ${\rm A}_1\to{\rm A}_2$ for $k_{+2}>0.6$.  The efficiency of these conversions is very small, because the contribution of the cycle $c=1$ to the entropy production rate is the smallest among the seven cycles, as seen in Table~\ref{table}.  Nevertheless, the study shows the possibility of changing the direction of chemical conversion even if the reaction network is running in chaotic regimes.


\section{Conclusion}
\label{Conclusion}

In this paper, methods to evaluate the entropy production rate in chemical chaos are developed in both the deterministic and stochastic approaches.  Since trajectories are time dependent, time average is considered in both approaches.  In general, time average defines some stationary probability distribution.  In deterministic systems, this stationary distribution is a solution of the generalized Liouvillian equation and it has the chaotic attractor for support.\cite{N95}  In stochastic processes, the stationary distribution is a solution of the chemical master equation.  The latter distribution is expected to reduce to the former in the weak noise limit.

The model of chemical chaos that is here considered\cite{GN83} is a reversible chemical reaction network composed of ten forward and so many backward reactions, ruling the time evolution of three intermediate species.  The bifurcation diagram of the deterministic dynamical system shows that the parametric domains with chaotic attractors are alternating with small windows of mixed-mode periodic oscillations.  In the corresponding stochastic process, the structures in phase space and parametric space that are finer than the noise amplitude are thus blurred by the random fluctuations.  However, the larger structures are observed to be robust.

In deterministic systems, it is shown for chemical kinetics obeying the mass action law that the time average of the entropy production rate can be decomposed in terms of the affinities and mean rates associated with the right null eigenvectors of the stoichiometric matrix.  These null eigenvectors define cycles composed of consecutive reactions such that the molecular numbers of intermediate species come back to their initial values.  The associated affinities only depend on the concentrations of the chemostatted species and the mean rates give the amounts of chemostatted species that are consumed or synthesized in the cycle.  The results of this paper show that this decomposition, which is well known in stationary states of deterministic chemical reaction networks, is also valid in time-dependent periodic, quasiperiodic, or chaotic regimes for the time average of the entropy production rate.

In this way, the comparison can be made with the entropy production rate evaluated for the corresponding stochastic process by averaging either in time or with respect to the stationary probability distribution of the process under the assumption of ergodicity.  In the stochastic approach, as well as for deterministic systems, the entropy production rate is evaluated using the rates of the forward and backward elementary chemical reactions.  For stochastic processes, the rates can be obtained by counting the random numbers of reactive events.  Every reactive event contributes to entropy production by the logarithm of the ratio of the opposite transition rates of the elementary reaction occurring at the transition.  Consequently, the known expression for the entropy production rate is recovered in the deterministic limit.  

Equivalently, the entropy production rate can be evaluated in the stochastic approach by counting the random numbers of cycles that are wound during the process, every cycle adding a contribution equal to the associated affinity.  The advantage of this method is that the entropy production rate can be evaluated as some linear combination of reaction rates.

The application of these methods to the chemical reaction network of Ref.~\onlinecite{GN83} in the stationary, periodic, and chaotic regimes shows that they are consistent with each other.  On the one hand, the values of the entropy production rate given by the cycle decomposition are in excellent agreement with the values of the standard method up to great numerical accuracy.  On the other hand, the values obtained for the stochastic process are observed to agree with the values of the deterministic system in the weak noise limit.

Furthermore, the cycle decomposition reveals that, in the chaotic regime, some cycles contribute negatively to the positive entropy production rate, meaning that these cycles are driven by the other ones in the direction that is opposite to spontaneous dissipation.  Accordingly, the chemical reaction network performs energy conversion even in the chaotic regime.  Moreover, the direction of conversion in one of the cycles can be reversed by tuning the associated control parameter, although remaining in the chaotic regime.  This cycle is driven opposite to its spontaneous direction below a critical value of this control parameter, where the thermodynamic efficiency of energy conversion is evaluated.

To conclude, the methods of stochastic thermodynamics can be applied to complex chemical reaction networks running from stationary to chaotic regimes and they are consistent with the standard methods of chemical nonequilibrium thermodynamics.


\begin{acknowledgments}
This research is financially supported by the Universit\'e libre de Bruxelles (ULB) and the Fonds de la Recherche Scientifique~-~FNRS under the Grant PDR~T.0094.16 for the project ``SYMSTATPHYS".
\end{acknowledgments}


\end{document}